\documentclass{jkas}

\def\year{2023} 
\def\month{June} 
\def\volume{00} 
\def\issue{0} 
\def\beginpage{1} 
\def\endpage{99} 
\def\received{June 08, 2023} 
\def\accepted{August 00, 2023} 
\date{Received \received; accepted \accepted}

\title{A Search for Exoplanets around Northern Circumpolar Stars VIII. Filter out a planet cycle from the multi-period radial velocity variations in M giant HD~36384
}

\author[1,2]{Byeong-Cheol Lee}
\author[3]{Gwanghui Jeong}
\author[4]{Jae-Rim Koo}
\author[4]{Beomdu Lim}
\author[5]{Myeong-Gu Park}
\author[5]{Tae-Yang Bang}
\author[1,2]{Yeon-Ho Choi}
\author[1]{Hyeong-Ill Oh}
\author[1]{Inwoo Han}

\affil[1]{Korea Astronomy and Space Science Institute, 776, Daedeokdae-Ro, Youseong-Gu, Daejeon 34055, Korea; \email{bclee@kasi.re.kr}}
\affil[2]{Korea University of Science and Technology, 217, Gajeong-ro Yuseong-gu, Daejeon 34113, Korea}
\affil[3]{Space Science Research Center, Antbridge Inc, 101, 79-1 Gajeong-ro, Yuseong-gu, Daejeon 34120, Korea}
\affil[4]{Department of Earth Science Education, Kongju National University, 56 Gongjudaehak-ro, Gongju-si, Chungcheongnam-do 314701, Korea}
\affil[5]{Department of Astronomy and Atmospheric Sciences, Kyungpook National University, Daegu 702-701, Korea}

\def\corrauthor{
Byeong-Cheol Lee
}

\def\runningauthor{
Byeong-Cheol Lee et al.
}

\def\runningtitle{
A Search for Exoplanets around Northern Circumpolar Stars VIII.
}

\def\keywords{
stars: individual: \mbox{HD 36384}, techniques: radial velocities, stars: planetary systems
}

\def\abstracttext{
This paper is written as a follow-up observations to reinterpret the radial velocity (RV)  of HD~36384, where the existence of planetary systems is known to be ambiguous.
In giants, it is, in general, difficult to distinguish the signals of planetary companions from those of stellar activities. Thus, known exoplanetary giant hosts are relatively rare. We, for many years, have obtained RV data in evolved stars using  the high-resolution, fiber-fed Bohyunsan Observatory Echelle Spectrograph (BOES)  at the Bohyunsan Optical Astronomy Observatory (BOAO).
Here, we report the results of RV variations in the M giant \mbox{HD 36384}.
We have found two significant periods of 586d and 490d. Considering the orbital stability, it is impossible to have two planets at so close orbits. To determine the nature of the RV variability variations,
we analyze the \textit{HIPPARCOS} photometric data, some indicators of stellar activities, and line profiles. A significant period of 580d was revealed in the \textit{HIPPARCOS} photometry. H$_{\alpha}$ EW variations also show a meaningful period of 582d.
Thus, the period of 586d may be closely related to the rotational modulations and/or stellar pulsations. On the other hand, the other significant period of 490d is interpreted as the result of the orbiting companion. Our orbital fit suggests that the companion was a planetary mass of 6.6 $M_{J}$ and is located at 1.3 AU from the host.
}

\begin{document}
\jkashead 


\section{Introduction\label{sec:intro}}
\begin{table*}
\renewcommand{\thetable}{\arabic{table}}
\centering
\caption{RV measurements for HD 36384 from December 2010 to February 2022 using the BOES.} \label{tab:rv2}
\begin{tabular}{ccccccccc}
\hline
\hline
JD & RV  & $\pm \sigma$ & JD & RV  & $\pm \sigma$  & JD & RV  & $\pm \sigma$\\
$-$2,450,000 &{m\,s$^{-1}$}& {m\,s$^{-1}$}& $-$2,450,000 & {m\,s$^{-1}$} & {m\,s$^{-1}$} & $-$2,450,000 & {m\,s$^{-1}$} & {m\,s$^{-1}$}\\
\hline
5554.160778 &     141.5   &    12.7 &  6617.039120  &    173.0  &     10.3 &  8562.032879 &    -210.1 &      10.5  \\
5843.296304 &    -134.0   &     9.2 &  6739.981866  &   -177.5  &     11.0 &  8829.101603 &      69.2 &       9.6  \\
5934.083138 &     348.0   &    10.2 &  6922.249499  &   -263.8  &      9.9 &  8862.049446 &     207.0 &      10.3  \\
5963.013631 &     317.8   &     8.9 &  6965.948245  &     55.3  &      9.3 &  8863.145400 &     266.5 &      12.4  \\
6015.070065 &     223.9   &    10.3 &  7094.044613  &    163.4  &     14.7 &  8945.032492 &     275.7 &      10.1  \\
6024.942705 &     305.9   &    11.6 &  7301.127580  &   -231.8  &     10.7 &  9511.130388 &     202.6 &      36.1  \\
6073.967887 &       6.0   &    13.7 &  7330.305666  &   -257.3  &      9.1 &  9511.130388 &     211.3 &      24.8  \\
6258.186458 &    -297.6   &    10.0 &  7378.120219  &   -248.4  &     11.3 &  9512.090507 &      78.7 &      23.6  \\
6261.212614 &    -385.8   &     9.0 &  7401.943197  &   -144.9  &     11.7 &  9512.092243 &      91.7 &      23.2  \\
6287.150081 &    -380.0   &    17.7 &  7407.925704  &   -126.3  &     10.3 &  9512.133912 &     111.9 &      27.4  \\
6346.073735 &    -188.5   &     9.2 &  7412.941397  &   -213.2  &     10.8 &  9633.050081 &    -247.1 &      28.1  \\
6378.102166 &     -60.8   &    11.0 &  7423.961470  &    -10.4  &     11.4 &  9634.008360 &    -176.7 &      25.3  \\
6551.314505 &     231.6   &    11.4 &  7469.046414  &     28.0  &      8.9 &  9635.008303 &    -212.9 &      39.5  \\
6582.262129 &     281.3   &    11.0 &  8515.956277  &     31.8  &     10.4 &  9636.008245 &     -95.0 &      22.1  \\
6583.166785 &     238.6   &    12.2 &               &           &          &              &           &            \\
\hline
\end{tabular}
\end{table*}

Even if the periodic radial velocity (RV) variations originate from the wobbling of a star by the orbital motions of planets, subsequent follow-up observations often obtain some conflict results from the previous observations.
Thus, long-term follow-up observations are critical to definitely identify the cause of the RV variation.
There are several cases related to such conflict results between previous and subsequent observations.

Some periodic changes appear to be associated with stellar activity.
\cite{2018AJ....155..120H} reported that the multi-periodic behavior of $\gamma$ Dra may represent a new form of stellar variability, possibly related to oscillatory convective modes.
They present precise stellar RV outcomes on $\gamma$ Dra taken from 2003 to 2017. The data from 2003 to 2011 show coherent, long-lived variations with a period of 702 days. These variations are consistent with the presence of an a actural planetary companion.
However, RV measurements taken from 2011–2017 seem to refute this.
They suggest that this  behavior may represent possibly related to oscillatory convective mode.
\cite{2015A&A...580A..31H} found RV signals of planetary companion and stellar activity around the K5 giant Aldebaran.
As the best Keplerian fit to the combined RV data, they provide further evidence of a planetary companion to the star as well as stellar activity variations.
However, the follow-up study of \cite{2019A&A...625A..22R} found no conclusive evidence for the existence of the planet.

The larger the sample, the higher the probability of finding other signals from other planets or stellar origins.
Although the spectral types are different, the observations of $\tau$  Ceti (G8 V) would be a good example.
Since $\tau$ Ceti is known to be an inactive star, this has been a good target for planet searching.
A couple of previous studies \citep{2006AJ....132..177W, 2011A&A...534A..58P} have reported that $\tau$ Ceti does not have any planetary objects around it, while another studies \citep{2013A&A...551A..79T, 2017AJ....154..135F} suggested the presence of multiple planets. Also, several similar cases \textbf{have} steadily reported., e.g., five planets and one candidate around HD 158259 \citep{2020A&A...636L...6H}, six planets
around GJ 667C \citep{2013A&A...553A...8D}, six planets around TOI 178 \citep{2021A&A...649A..26L}, and four
planets around HD~20781 \citep{2019A&A...622A..37U}.

Since 2003, we have been searching for exoplanets using the RV technique based on the homogeneous data sets from the same instrument. The  star HD 36384 shows a long-period  RV variation \citep{2017ApJ...844...36L}, however, its origin is rather unclear. Therefore, further data were required to arrive at a definite conclusion on the origin of the variation. In this study, we revisit HD 36384 to uncover the origin of the RV variation using a larger set of RV data. In Section 2, our observations and stellar parameters of this star are addressed. We analyze the periodogram in Section 3. The signals from stellar activities are discussed in Section 4. In Section 5, we discuss the overall results obtained from this study.

%
\begin{table*}
\begin{center}
\caption{Stellar parameters for HD 36384.}
\label{tab:ste}
\begin{tabular}{lcc}
\hline
\hline
	Parameter					& Value	    & Reference\\
\hline
Spectral type					& M0 III	&\textit{HIPPARCOS} \citep{1997yCat.1239....0E}\\
$\textit{$m_{v}$}$ (mag)			& 6.19	& \textit{HIPPARCOS} \citep{1997yCat.1239....0E}\\
$\textit{B -- V}$ (mag)			& 1.606  	& \textit{HIPPARCOS} \citep{1997yCat.1239....0E}\\
${HIP_{scat}}$ (mag)			& 0.012	    & \textit{HIPPARCOS} \citep{1997yCat.1239....0E}\\
$\pi$ (mas)					& 5.10 $\pm$ 0.58   & Gaia Collaboration et al. (2018)\\
$T_{\rm{eff}}$ (K)				& 3940 $\pm$ 40	& \cite{2017ApJ...844...36L}\\
$\rm{[Fe/H]}$				& $-$0.16 $\pm$ 0.14& \cite{2017ApJ...844...36L}\\
log $\it g$ (cgs)				& 1.1 $\pm$ 0.2	& \cite{2017ApJ...844...36L}\\
$v_{\rm{micro}}$ (km s$^{-1}$)	& 6.8 $\pm$ 2.7	& \cite{2017ApJ...844...36L}\\
Age (Gyr)						& 0.113 $\pm$ 0.015      & \cite{2012AstL...38..331A}\\
$\textit{$R_{\star}$}$ ($R_{\odot}$)	& 38.4 $\pm$ 3.4 & \cite{2017ApJ...844...36L}\\
$\textit{$M_{\star}$}$ ($M_{\odot}$)  & 1.14 $\pm$ 0.15  & \cite{2017ApJ...844...36L}\\
$\textit{$L_{\star}$}$ ($L_{\odot}$)  & 388.28 $\pm$ 0.15& \cite{2017MNRAS.471..770M} \\
$v_{\rm{rot}}$ sin $i$ (km s$^{-1}$)  & 4.5 $\pm$ 0.1	 & \cite{2017ApJ...844...36L}\\
$P_{\rm{rot}}$ / sin $i$ (days)		  & 440 $\pm$ 40	 & \cite{2017ApJ...844...36L} \\
\hline
\end{tabular}
\end{center}
\end{table*}

\section{Observations, Reduction, and Stellar Characteristics\label{sec:star}}
We obtained 43 spectra in total for \mbox{HD 36384} during the period of 2010 to 2022 with the BOES  mounted on the 1.8 m telescope at the Bohyunsan Optical Astronomy Observatory (BOAO) in Korea. The BOES spectrograph covers wavelength between 3,500\AA\ and 10,500\AA\ and provides a resolving power of $R$ = 90,000. 
To minimize any measurement errors due to spectral line broadening (e.g., long-term exposure),
 an exposure time was restricted to 15 minutes and the signal-to-noise ratio at the iodine (I$_{2}$) absorption spectrum reached approximately 150.
The IRAF package was used to process the raw data. The RV standard star $\tau$ Ceti has been monitored since 2003, showing rms scatter of about \mbox{7 m s$^{-1}$} \mbox{\citep{2013A&A...549A...2L}}.
The basic stellar parameters were based on the \textit{HIPPARCOS} catalog \citep{1997yCat.1239....0E} and values by \cite{2017ApJ...844...36L}.
The stellar parameters for \mbox{HD 36384} are listed in Table 2.


\section{Orbital solutions}

In order to determine the periodicity in the BOES RV variations, we used the Generalized Lomb-Scargle periodogram \citep[GLS;][]{2009A&A...496..577Z}.
We determined the orbital elements by fitting the RV with Keplerian orbit model. The RV measurements and the RV curve for HD~36384 is shown in Table 1 and Fig. 1.
Measured precision RV data  revealed two large amplitude periodic signals of 586d and 490d [Figure 2 (a)].
We initially subtract the significant period of 586d, and then use an iterative prewhitening procedure, as described in \cite{2011A&A...533A...4B}, to characterize additional variability in the residuals for HD~36384. The second period appears at 490d [Figure 2 (b)].
However, not both periods are likely to be caused by planets, as a system with two planets this close is very mechanically unstable.
Therefore, in order to determine the actual cause,  several tests of the most likely period were conducted using the periodogram.
Figure 2 (c) shows the power spectrum of the velocity residuals after removing the two periodic signals, indicating that no significant variations exist. Figure 2 (d) shows a window function.

In this paper, we  investigated the line shape of the HD~36384 spectra again to check for RV variations with other origins.
Two bisector quantities were calculated from the line profile at two flux levels (40\% and 80\%) (the bisector velocity span BVS = V$_{top}$ -- V$_{bottom}$] and the bisector velocity curvature BVC = [V$_{top}$ -- V$_{center}$] -- [V$_{center}$ -- V$_{bottom}$].
To search for variations in the spectral line shape, we selected unblended spectral feature with high flux level leaved iodine absorption region: V1 6039.722~\AA.
We carried out GLS period analyses of the bisectors as shown in the Fig. 2 (f).
No significant peak is seen, as in earlier work \citep{2017ApJ...844...36L}.

In the second panel in Fig 2, another significant peak appears around 210d.
However, it appears to be a weak signal (FAP $\sim$ 10\%) in periodogram of the residual RVs after subtracting two strong peaks [Fig 2 (c)], thus, may be disregarded. We could not find third periodic signal in the RV residuals.

Regarding prominent two peaks at 586d and 490d  in the RV power spectrum, we computed a preliminary FAP lower than the 1 $\times 10^{-7}$\% level on the periodogram of the RVs. We then used RVI2CELL \citep{2007PKAS...22...75H} for the best fit to yield the orbital parameters. 
The bootstrap randomization method \citep{1999A&A...344L...5K} tool was used to calculate the uncertainty of the orbital parameters. We calculated the highest peak in 200,000 trials in this study.
The derived two orbital parameters of the \mbox{HD 36384} system are as an orbital period of 586 $\pm$ 4d, a semi-amplitude outcome of 206 $\pm$ 12 m s$^{-1}$, and eccentricity of 0.2 $\pm$ 0.1 and an orbital period of 490 $\pm$ 3d, a semi-amplitude outcome of 156 $\pm$ 14 m s$^{-1}$, and eccentricity of 0.2 $\pm$ 0.1.


\begin{figure}[h!]
\centering
\includegraphics[width=85mm]{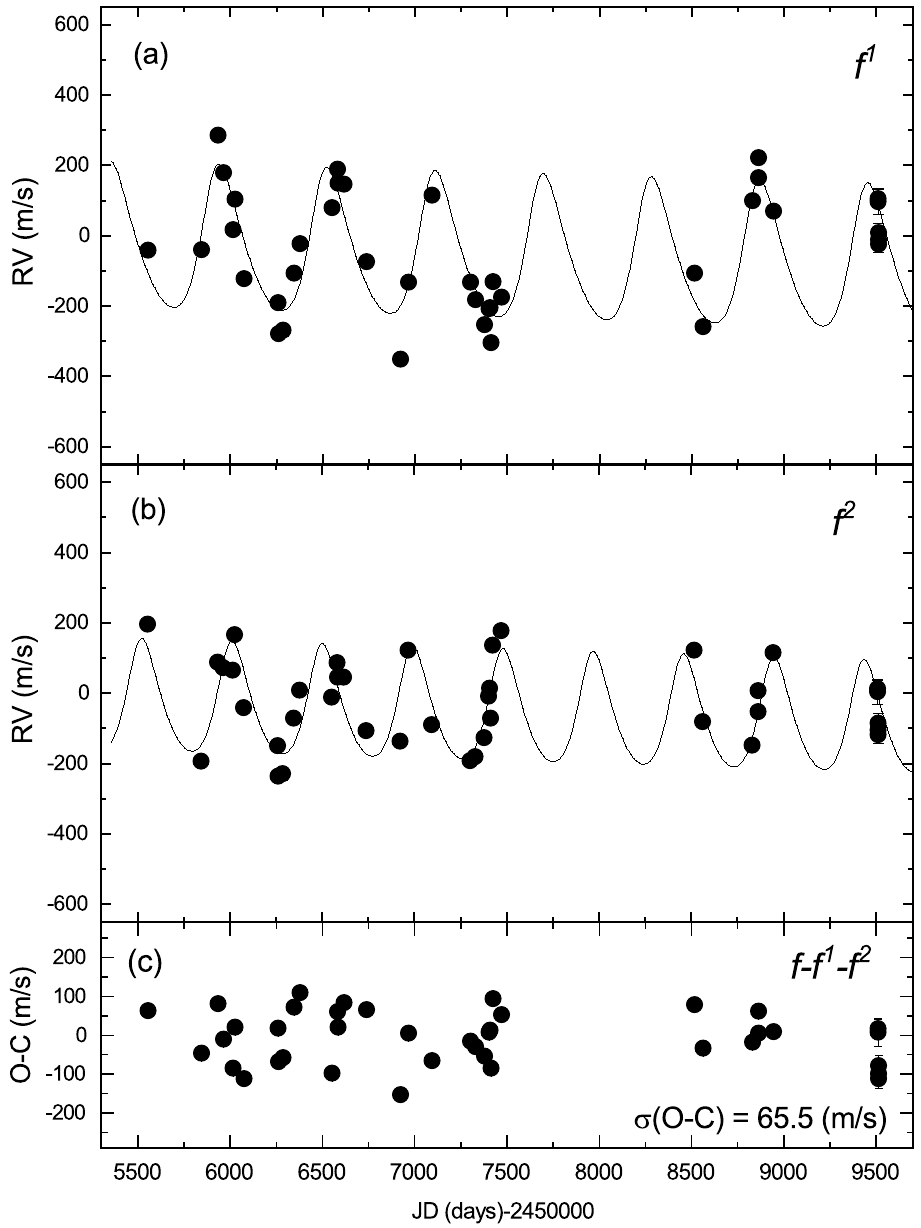}
\caption{RV curve for HD 36384 (a) The best Keplerian orbit model is shown as a solid line (b) RV curve after subtracting the first fit from the (a), and (c) Residual RVs after subtracting two strong orbits}. \label{fig:fig2}
\end{figure}

%


%

\begin{figure}[h!]
\centering
\includegraphics[width=100mm]{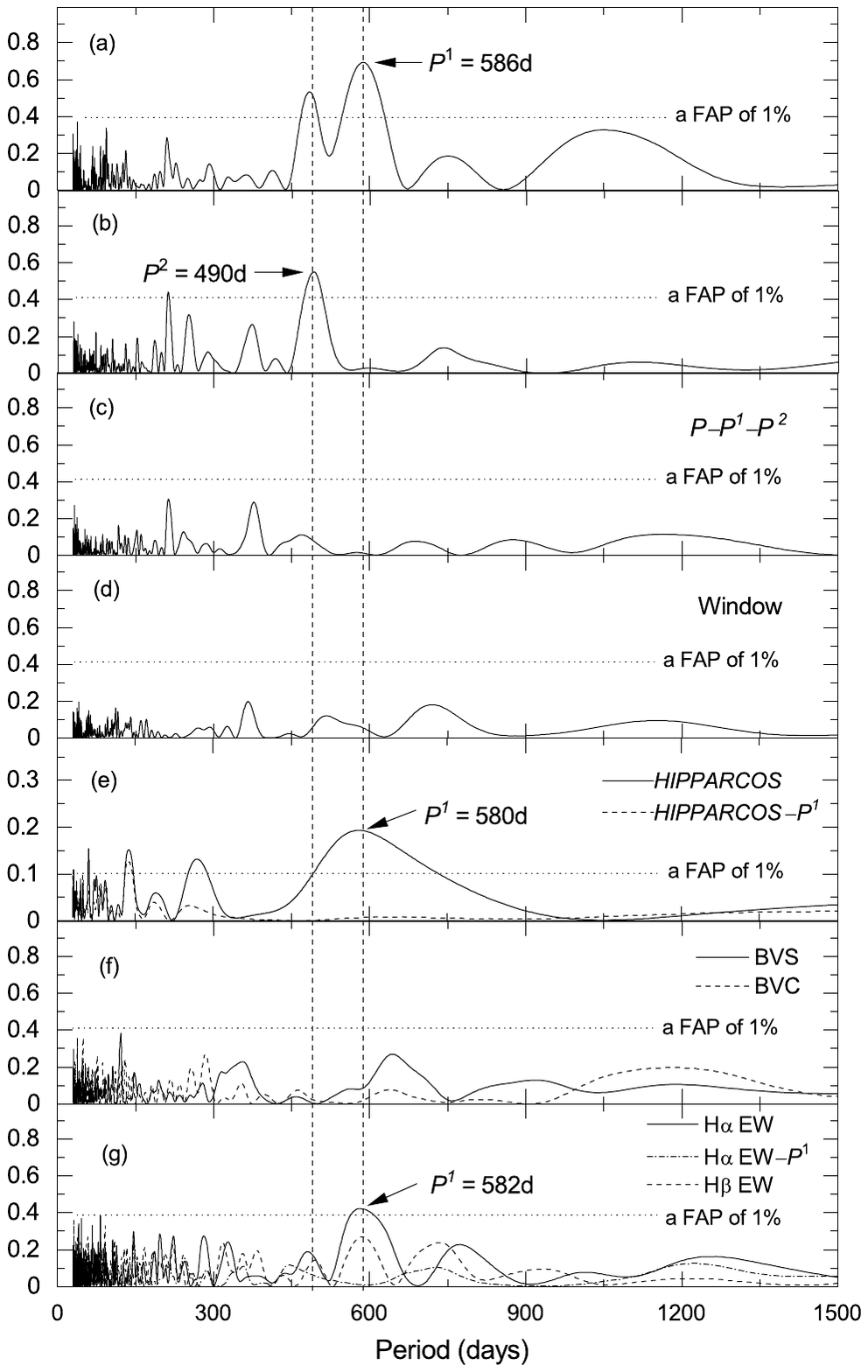}
\caption{(a) GLS periodogram for RVs of \mbox{HD 36384} (b) The same periodogram for the residual RVs after subtracting the strong peak of 586d (c) periodogram of the residual RVs after subtracting two strong peaks (d) Periodogram of our time sampling (window function) e) periodogram of the \textit{HIPPARCOS} photometric data (f) periodograms of the line bisector span (BVS) and the line bisector curvature (BVC), and (g) H line EW variations. The vertical dashed lines indicate orbital periods of 490d and 586d. The horizontal lines in each panel correspond to a 0.1\% FAP.
\label{fig2}}
\end{figure}

\section{Stellar Origin\label{sec:nature}}

The long period RV variations seen in HD~36384 can be originated from rotational modulation, pulsation, and/or orbiting companions.
To determine the origin of  these periodic RV variations, it is necessary to analyze additional diagnostic indicators such as surface activity (Ca~II~H line and H lines variations), bisector variations, and photometric variations.
We have obtained additional data, and reanalyzed these diagnostic indicators.

The Ca II line is sensitive to chromospheric activity, and manifests the activity though emission lines.
The Ca II line of HD~36384 shows no noticeable emission feature at the core of the line, as shown in Fig. 10 of \citet{2017ApJ...844...36L}.
We additionally used H lines, which are also useful when assessing chromospheric activity.
It is important to avoid nearby blending lines and weak telluric lines.
We measured the H line EWs using a band pass value of $\pm$ 1.0 $\rm\AA$ centered at the core of the H lines.
The rms in the H$_{\alpha}$ and the H$_{\beta}$ EWs correspond to less than 0.1\% variation.
The GLS periodograms of the H line EW variations are shown in Fig.~\ref{fig2} (g).
There is a significant power at the period of 582d, close to the period of 586d seen in RV variations.

M-type giants such as HD 36384 show changes in brightness mainly due to changes in the intensity of cool points.
The \textit{HIPPARCOS} photometric data were obtained between December of 1989 and March of 1993, and these represent 177 observations for \mbox{HD 36384}. The data has rms scatter \textbf{as large as }0.19 mag, indicating considerable scatter.
GLS periodogram analyses of the photometric data indicate a strong peak at 580d.
The total range of the photometric variations is greater than two tenths of the magnitude.

In general, giants reveal pulsation periods of a few days in the form of radial pulsation.
Using the scaling relation Eq. (7) from \cite{1995A&A...293...87K} with a luminosity and a mass of HD~36384, the radial pulsation period and the expected  amplitudes can be estimated (Table 3). 
The periods are far too short to be an explanation for the observed RV variations. The dispersion of the RV residuals is 67~m s$^{-1}$, which is significantly higher than the rms scatter of the RV standard star (7 m~s$^{-1}$) and the typical RV measurements error ($\sim$ 17 m~s$^{-1}$). We thus can propose that the large scattering of the residuals can be attributed to unresolved oscillations.


\begin{table}
\begin{center}
\caption{Radial pulsation modes for HD 36384.}
\label{tab6}
\begin{tabular}{lcc}
\hline
\hline
    Mode                              &               &  \\
\hline
    Fundamental period       & [days]        & 19    \\
    Pulsation period         & [days]        & 6   \\
    Pulsation RV amplitude   & [m s$^{-1}$]  & 80    \\
\hline
\end{tabular}
\end{center}
\end{table}
%

\begin{table}[h!]
\renewcommand{\thetable}{\arabic{table}}
\centering
\caption{Preliminary orbital solutions of HD 36384 b.} \label{tab:orb}
\begin{tabular}{lc}
\hline
\hline
    Parameter	& Value\\
\hline
P (days)					& 490 $\pm$ 3 \\
K (m s$^{-1}$)				& 156 $\pm$ 14 \\
$e$						& 0.2 $\pm$ 0.1 \\
$T_{periastron}$ (JD) 	& 2455982.8 $\pm$ 29.8 \\
$\omega$ (deg)			& 197.6 $\pm$ 34.6 \\
$m$ sin $i$ ($M_{J}$)			& 6.6 $\pm$ 0.5 \\
$a$ (AU)         				& 1.3 $\pm$ 0.1\\
$Slope$  (m s$^{-1} yr^{-1}$)   & 1.5	$\times$ 10$^{-5}$ \\
$N_{obs}$               & 43	 \\
rms (m s$^{-1}$)			    & 67 \\
\hline
\end{tabular}
\end{table}

\section{Discussion\label{sec:dis}}
Generally, most giant stars have intrinsic RV variations, pulsations and/or surface activities, which can produce RV variations similar to those from orbiting planets.
In order to determine the nature of the RV variabilities, we should undertake all relevant analyses comprehensively that can identify the real origin of the RV variations.

\cite{2017ApJ...844...36L} found a RV period of 535d in HD~36384, though their study did not show  any obvious evidence of RV variations caused by a rotational modulation of the surface features in the H line EW variations or line bisector measurements of HD~36384.
However, it was suspected that the \textit{HIPPARCOS} photometric period of 580d was related to the RV period.
Also, in order to detect transient periodic fluctuations or changes, \cite{2017ApJ...844...36L} conducted a weighted wavelet Z-transformation \citep[WWZ;][]{1996AJ....112.1709F} analysis.
The WWZ analysis provides clue with regard to a secular decrease in the amplitude of RV variations over long-time periods.  They showed that the maxima period scarcely changed over the observation periods. However, the amplitude of the RV variation decreased by more than 30\%, suggesting that RV variations are unlikely to be caused by a companion.
WWZ testing may not have been appropriate at the time because it did not take into account multiple cycles.
Thus, \cite{2017ApJ...844...36L} interpreted the RV variations in \mbox{HD 36384}, which as those from a pulsations rather than those by a planet.

Through follow-up and reinterpretation, we now find two significant periods of 586d and 490d, rather than a single period of 535d.
A prominent periodic signal of 586d is shown to be close to photometric measurements and the H$_{\alpha}$ line EWs.
First, it is necessary to check the correlation between the \textit{HIPPARCOS} photometric data and the period of the RV signal. 
In the discovered multi-mode from RV analysis, one may occur from the rotation modulations of the star and the other from a companion. The presence of large inhomogeneities of surface brightness is expected due to the surface rotational modulation by the produced spot. The GLS periodogram analysis for the \textit{HIPPARCOS} photometric data shows a significant peak at around 580d.
This coincides with the RV period of 586d and appears  to be due to rotational modulation. Second, the H$_{\alpha}$ EWs variations were found to be 582d close to the RV period. This also suggests the possibility of being affected by a pulsation.

Exploration of planets orbiting bright giants with relatively large radii, such as HD 36384, requires further experiments.  The possibility of the long secondary periods (LSP), which are believed to be caused by large-scale stellar spot activity due to irregularities in the chromatography and light curves of red giants, should be considered. (\cite{2004ApJ...604..800W}; \cite{2015MNRAS.452.3863S}). The velocity amplitude of the LSP variable studied so far is a few km s$^{-1}$ (\cite{2002AJ....123.1002H}; \cite{2004ApJ...604..800W}). The second component in the LSP stars is most likely a brown dwarf.  The period is 200 to 1500 days and has a V band amplitude of up to one magnitude. However, in the case of HD~36384, the amplitude is 156 m s$^{-1}$ and the variable light is only 0.19 magnitude, so there seems to be no association of LSP.

In addition, stars with large radii (or large luminosities) and long periods are highly challenging  targets for planetary exploration. \citet{2021ApJS..256...10D} performed sanity tests on planets around evolved stars. That is, they propose a radius-period plot as a tool to determine the validity of planetary companions around more demanding host stars to rule out intrinsic stellar variability. 
A giant star with a radius of less than 21 $R_{\odot}$ shows a wide range of orbital periods, while no planets have relatively short (300 days) and long (800 days) periods around larger stars. In other words, it is presumed to mean a new phenomenon occurring in stars with radii greater than $\sim$ 21 $R_{\odot}$. The period of M giant HD 36384 with a radius of 38.4 $R_{\odot}$ is 490 days, which corresponds to the proposed range (300 days $<$ P $<$ 800 days).

In summary, RV monitoring of HD 36384, which was discontinued in 2017, was resumed in 2019 to better investigate these long-term variations. After follow-up observations and new analysis, we found two significant peaks in the RV signals.
Of the two, one stemmed from rotation modulations and/or pulsations and the other is due to a planet.
The best Keplerian fit to HD~36384 yields the following orbital parameters: an orbital period of 490 $\pm$ 3 d, a semi-amplitude value of  156 $\pm$ 14 m s$^{-1}$, and eccentricity of  0.2 $\pm$ 0.1.
 By adopting a stellar mass of 1.14 $\pm$ 0.15 $M_{\odot}$ for \mbox{HD 36384}, we obtain a minimum companion mass of 6.6 $\pm$ 0.5 $M_{J}$ and a semi-major axis of 1.3 $\pm$ 0.1 AU.


\acknowledgments{
BCL acknowledges partial support by the KASI (Korea Astronomy and Space Science Institute) grant
2023-1-832-03 and acknowledge support by the National Research Foundation of Korea(NRF) grant funded by the Korea government(MSIT) (No.2021R1A2C1009501).
MGP was supported by the Basic Science Research Program through the National Research Foundation of Korea (NRF) funded by the Ministry of Education (2019R1I1A3A02062242) and KASI under the R\&D program supervised by the Ministry of Science, ICT and Future Planning.
BL and JRK  acknowledge support by the NRF grant funded by MSIT (Grant No. 2022R1C1C2004102)
This research made use of the SIMBAD database, operated at the CDS, Strasbourg, France.
}



\end{document}